\documentclass[epjCONF]{svjour}
\pdfoutput=1
\usepackage{graphicx}
\usepackage[varg]{txfonts} 
\usepackage[latin1]{inputenc}
\usepackage{textcomp}

\session-title{Resonance Workshop at UT Austin}
\begin{document}
\title{Dielectron Measurements in STAR}
\author{Frank Geurts\fnmsep\thanks{\email{geurts@rice.edu}} (for the STAR Collaboration)}
\institute{Rice University, Houston TX 77005}

\abstract{Ultrarelativistic heavy-ion collisions provide a unique environment to study the properties of strongly interacting matter. Dileptons, which are not affected by the strong interactions,  are an ideal penetrating probe. We present the dielectron results for p$+$p and Au$+$Au collisions at $\sqrt{s_\mathrm{NN}}=$200~GeV, as measured by the STAR experiment. We discuss the prospects of dilepton measurements  with the near-future detector upgrades, and the recent lower beam energy Au$+$Au measurements.}

\maketitle

\section{Introduction}
\label{intro}
 Experimental results from heavy-ion collisions at the Relativistic Heavy Ion Collider (RHIC) have established that strongly-interacting hot QCD matter is created in Au+Au collisions at $\sqrt{s_{\rm NN}}=200$GeV \cite{whitepaperBRAHMS,whitepaperPHOBOS,whitepaperSTAR,whitepaperPHENIX}.  Leptons have a very low cross section with the strongly interacting medium. Dilepton pairs are created throughout the evolution of the strongly interacting system with its sources varying as a function of the kinematics. Thus, dileptons provide an excellent penetrating probe of the evolution of a heavy-ion collision. 
 
  The dilepton invariant mass spectrum can typically be divided into three regions, each defined by its dominant sources. In the lower mass region (LMR;   $m_\mathrm{ll} < 1.1$~GeV/$c^2$) dilepton pairs originate from the decay of vector mesons, such as the $\rho(770)$, $\omega(782)$, and $\phi(1020)$. Here, the $\rho$ meson is of special interest  given its relatively short lifetime of 1.3~fm/c, which is less than the expected lifetime of the system. This provides an opportunity to study the in-medium properties of such a vector meson, and relate any modifications of its spectral shape to a possible chiral symmetry restoration. Other LMR sources include the Dalitz decays of the pseudoscaler mesons, such as the $\pi^0$, $\eta$, and $\eta'(958)$. 

In the intermediate mass range (IMR; $1.1 < m_\mathrm{ll} < 3~$GeV/$c^2$), the production of dilepton pairs is closely related to the thermal radiation from the Quark-Gluon Plasma (QGP). However, significant backgrounds exist, including the heavy-flavor leptonic decays.

In the high mass region (HMR;  $m_\mathrm{ll} > 3$~GeV/$c^2$), the dilepton pairs are produced in the initial hard collisions between the partons of the incoming nuclei, as described in the Drell-Yan processes. In addition to these processes, another source of dilepton production in this range is the semileptonic decays of heavy quarks such as charm and bottom. These heavy quarks have been produced in the initial hard collisions. Dileptons in this region thus probe the initial stages of the heavy-ion collision. Furthermore, in this mass range, the dileptons result from the decay of heavy quarkonia, such as the J/$\psi$ and $\Upsilon$ meson. Yield measurements of these charmonium and bottomonium states provide an ideal opportunity to study deconfinement effects in the hot and dense medium.

Elliptic flow measurements are used to characterize the azimuthal asymmetry of momentum distributions. Dilepton elliptic flow measurements as a function of transverse momentum $p_\mathrm{T}$ have been proposed as an independent measure to study the medium properties \cite{Chatterjee}. The quantitative measure, $v_2$, is the second order harmonic of the azimuthal distributions with respect to the collision reaction plane.  In this paper we show the first preliminary results of the integral elliptic flow of the LMR and IMR dielectrons. For the mass ranges of the two pseudoscaler mesons, $\pi^0$ and $\eta$, we will also show the transverse momentum dependence of $v_2$ and compare to other elliptic flow measurements.

In 2010, the Solenoidal Tracker at RHIC (STAR) completed its installation of the Time-of-Flight (TOF) detector \cite{llope}. The TOF detector brings large-acceptance and high-efficiency particle identification that not only extends the hadron identification reach to higher momenta, but also significantly improves the electron identification in the low-momentum range. In this paper, we present the dielectron invariant mass spectra, and transverse mass distributions, for p$+$p and Au$+$Au collisions at $\sqrt{s_\mathrm{NN}}=$200~GeV. For the Au$+$Au system, we report the dielecton elliptic flow measurements.  Finally, we discuss the prospect of dielectron measurements at lower beam energy Au$+$Au collisions, and the near future upgrade plans for STAR which includes the Muon Telescope Detector (MTD) \cite{mtd}.

\section{Electron Identification and Background Reconstruction}
\begin{figure}[h]
\centering
\includegraphics[width=0.43\textwidth,keepaspectratio]{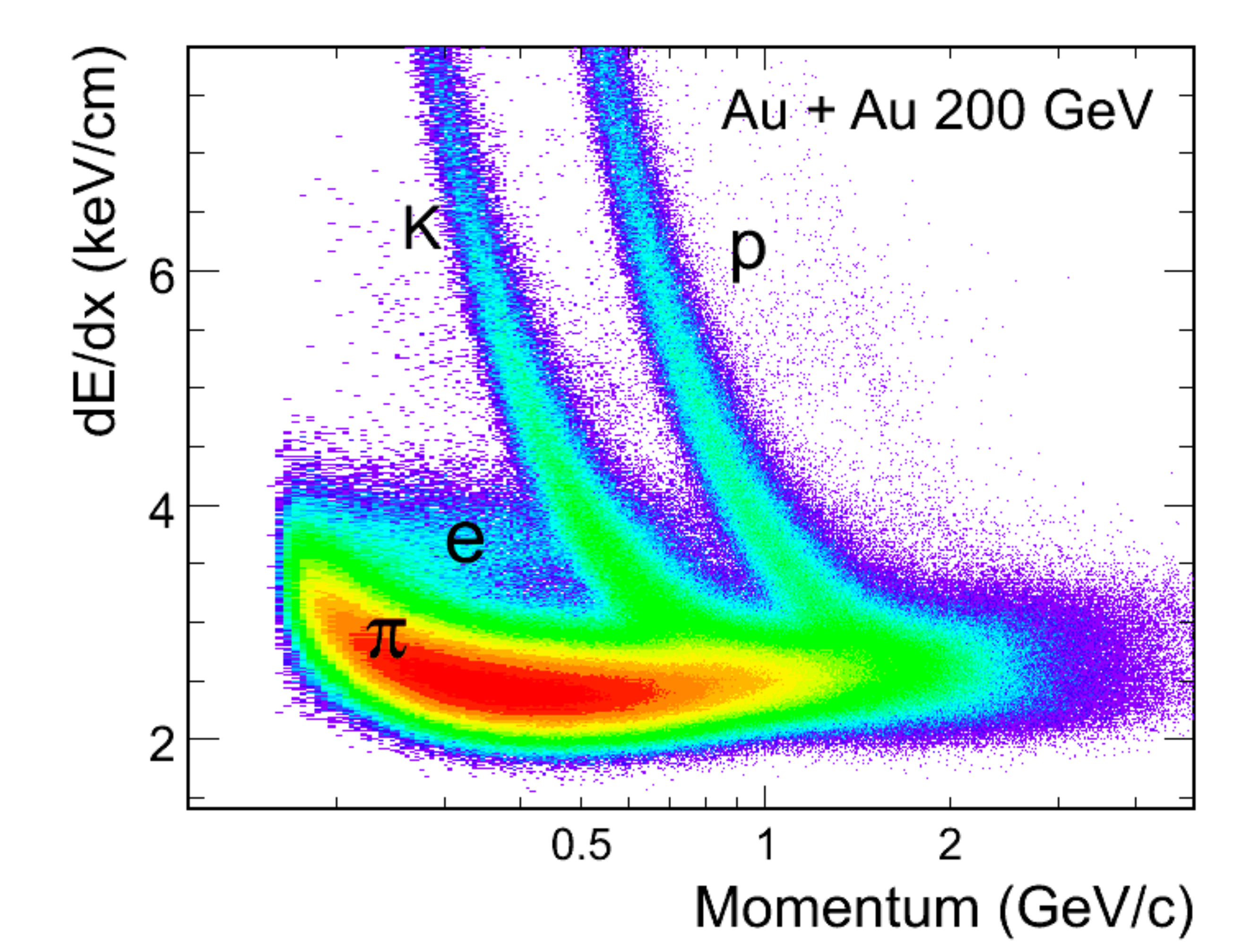}
\includegraphics[width=0.45\textwidth,keepaspectratio]{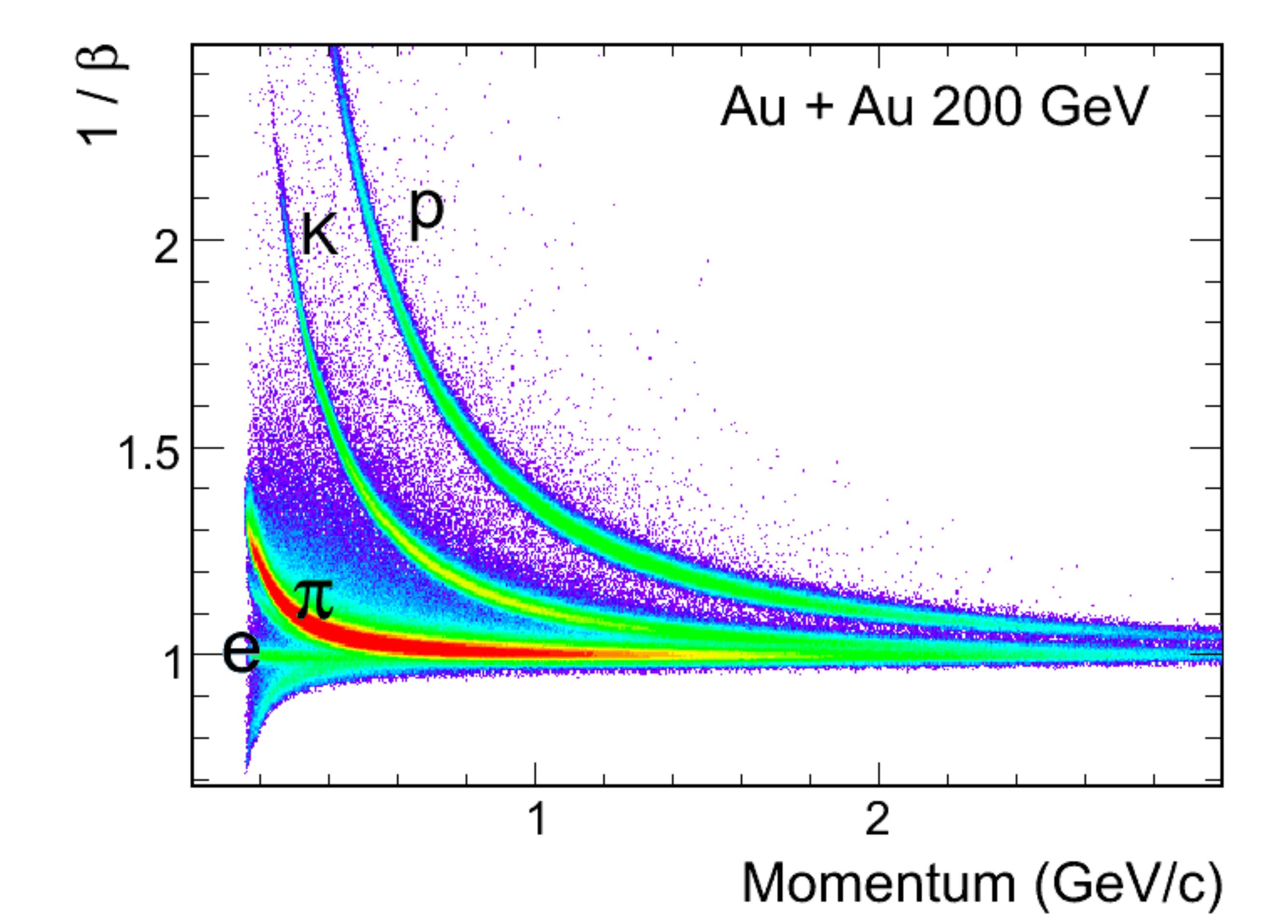}
\includegraphics[width=0.45\textwidth,keepaspectratio]{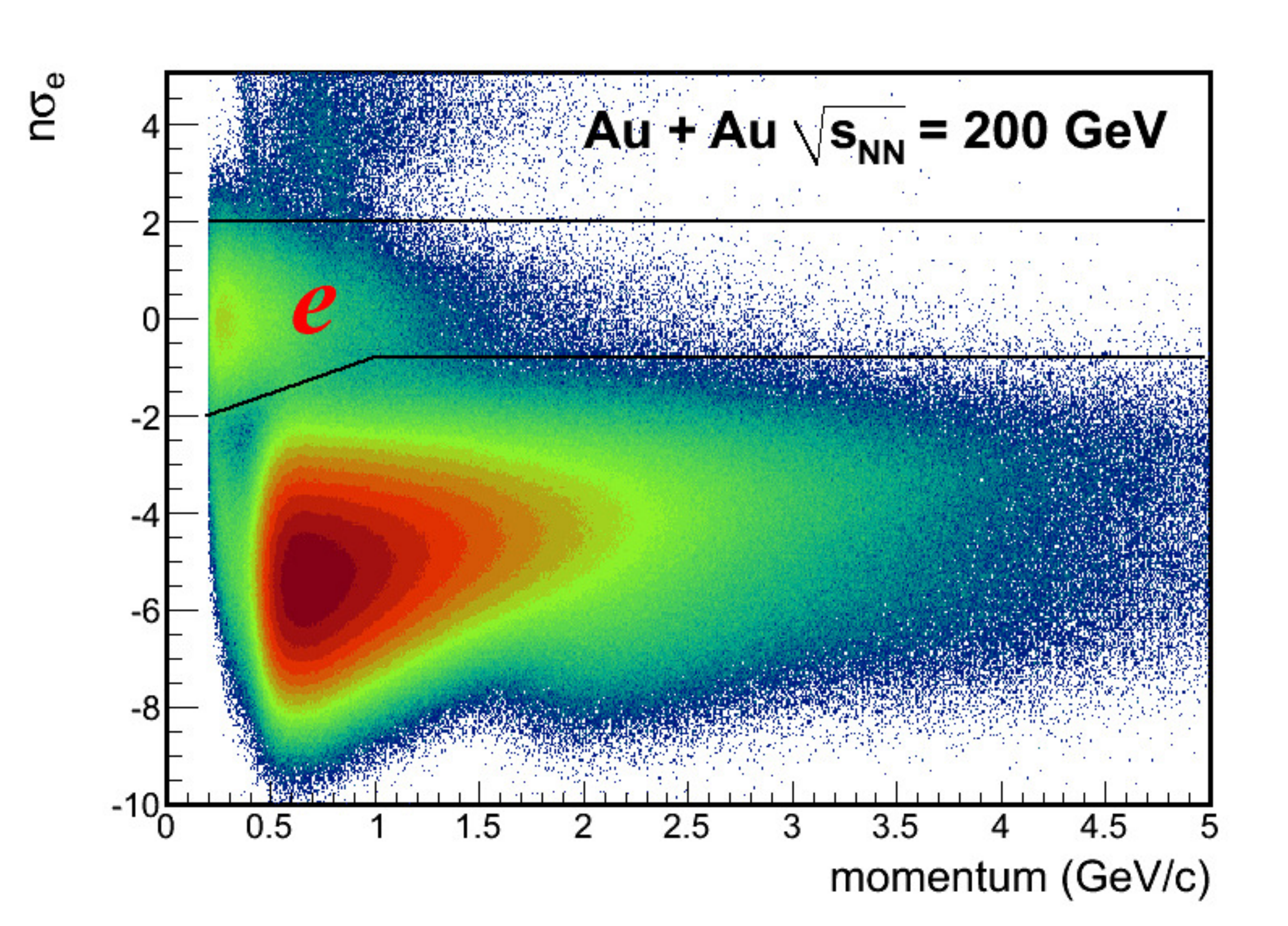}
\caption{Particle identification in the STAR experiment. Upper left panel: energy-loss, d$E$$/$d$x$, measurements in the TPC detector. Upper right panel: time-of-flight, $1/\beta$, measurements in the TOF detector. Lower panel: improved electron identification by combining the TOF and TPC measurements.}
\label{fig:epid}
\end{figure}

The electron identification for the results reported in the next section, involve the STAR Time Projection Chamber (TPC) and the TOF detector.  The TPC detector is the central tracking device of the STAR experiment and allows the reconstruction of particle tracks and momenta. The energy-loss measurements, dE$/$d$x$, in the TPC are used for particle identification. However, as can be seen in the upper-left panel of Fig.\ \ref{fig:epid}, identified particle bands merge with increasing momenta $p$ at relatively low values of $p$.
The TOF detector, with a 2$\pi$ azimuthal coverage at mid-rapidity, extends the particle identification range to higher momenta, as can be seen in the upper-right panel of Fig.\ \ref{fig:epid}. The velocity information distribution, $1/\beta$, as a function of momentum shows the electron band merge at low momenta with $\pi$ mesons. On the other hand, the electron band in the  dE$/$d$x$ distribution suffers from contamination by the intersecting hadron bands.
 By requiring  $|1-1/\beta|<0.025$ (0.03 in p$+$p collisions), the combination of the TOF velocity information and the TPC energy loss allows the removal of slower hadrons. In the lower panel of Fig. \ref{fig:epid} the resulting $n(\sigma_\mathrm{e})$ distribution is shown, as well as the selection window that is used in the Au$+$Au analysis. The value of $n(\sigma_\mathrm{e})$ is based on the log of the ratio of the measured and the expected average energy loss of electrons in the TPC
\[ 
 n(\sigma_\mathrm{e}) = \frac{1}{\sigma_\mathrm{e}}\log\frac{(\mathrm{d}E/\mathrm{d}x)^\mathrm{measured}}{\langle\mathrm{d}E/\mathrm{d}x \rangle^\mathrm{electron}},
\]
where $\sigma_\mathrm{e}$ is the pathlength-dependent energy-loss resolution for the electrons. With this selection window and a single track momentum threshold of $p_\mathrm{T}>0.2$~GeV/$c$, the electron purity in the p$+$p analysis is 99\% and in minimum bias Au$+$Au 97\%.

The invariant mass distributions are reconstructed by combining electrons and positrons from the same event. These so-called unlike-sign distributions have both signal and background contributions. Especially in high-multiplicity events, the contribution of the combinatorial background is substantial. The analyses that are presented in this paper use two different methods to determine the background. Each method has advantages and disadvantages that will be briefly discussed.

\begin{figure}[h]
\centering
\includegraphics[width=0.45\textwidth,keepaspectratio]{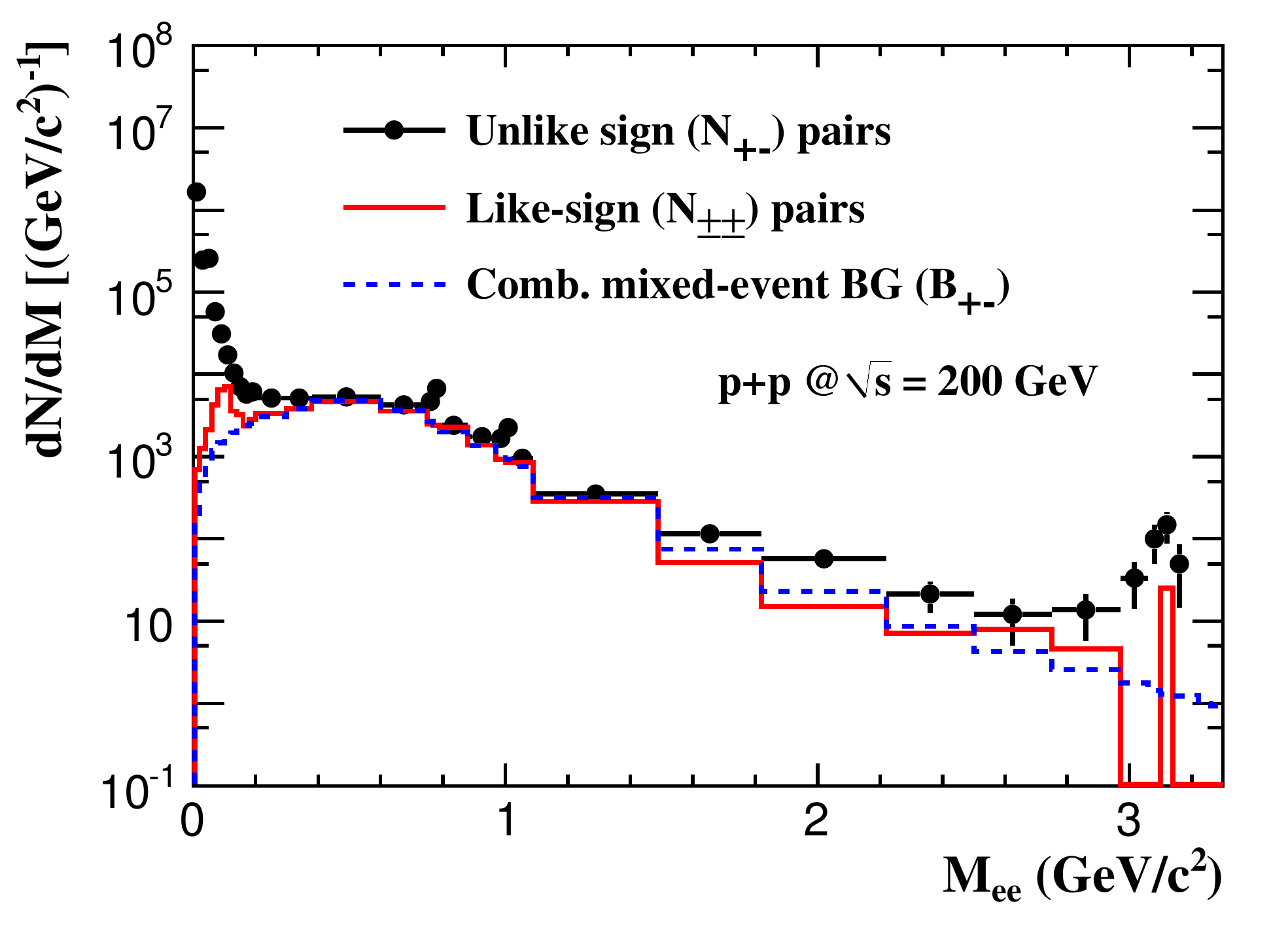}
\includegraphics[width=0.45\textwidth,keepaspectratio]{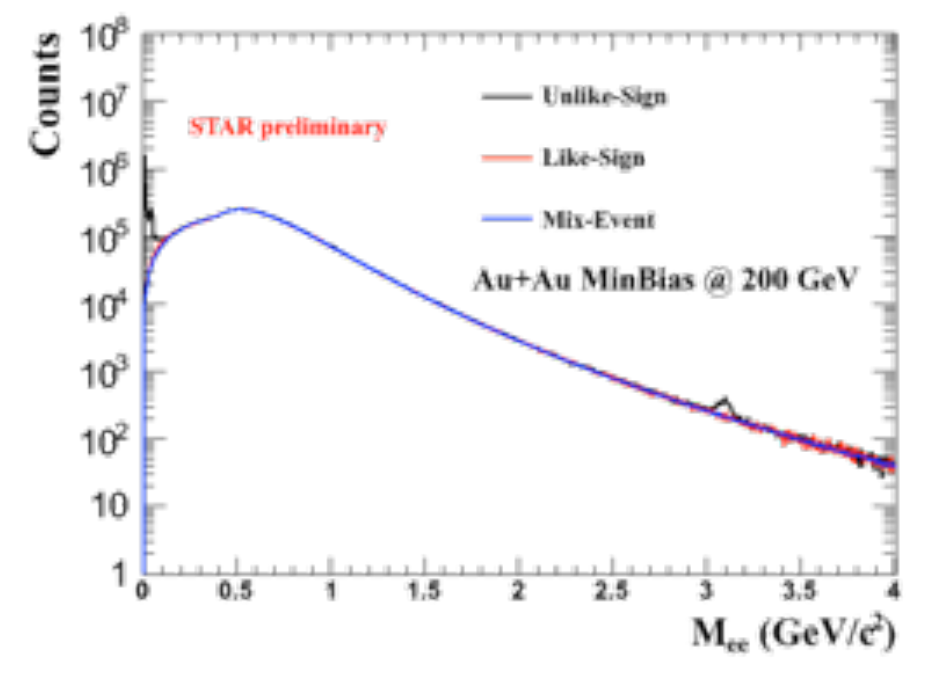}
\caption{Unlike-sign, like-sign, and mixed-event distributions in p$+$p \protect\cite{starpp} (left) and Au$+$Au minimum-bias events \protect\cite{Jie} (right) at $\sqrt{s_\mathrm{NN}}=200$~GeV.}
\label{fig:backgrounds}
\end{figure}

The mixed-event method combines electrons and positrons from different events. The events are categorized according to several characteristics such as the total particle multiplicity, event vertex location, and in the case of Au$+$Au collisions, the event plane angle. While the statistical accuracy of the background description can arbitrarily be improved by involving more events, the mixed-event method fails to reconstruct correlated background sources. At the lower invariant masses, such correlated pairs arise from jets, double Dalitz decays, Dalitz decays followed by a conversion of the decay photon, or two-photon decays followed by the conversion of both photons \cite{Lijuan}. A background determination in which like-sign pairs of the same event are combined is able to account for such correlated contributions. Its drawback, however, is that the statistical accuracy of this like-sign method is only comparable to the unlike-sign, {\em i.e.} original raw mass spectrum. Moreover, the like-sign method is more sensitive to detector acceptance differences than the mixed-event methods when compared to the unlike-sign spectrum.

\begin{figure}[ht]
\centering
\includegraphics[width=0.5\textwidth,keepaspectratio]{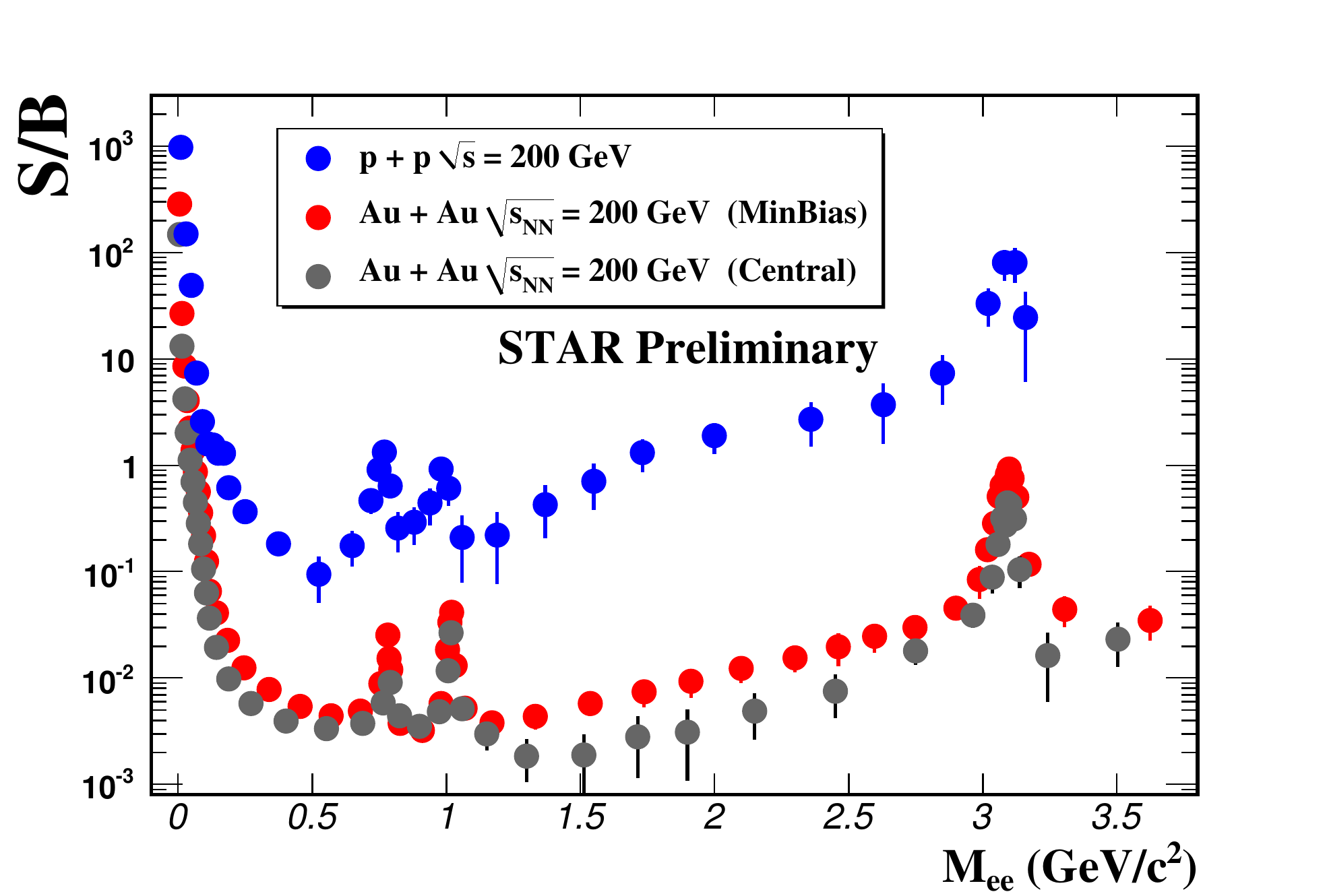}
\caption{Signal-to-background ratios in p$+$p, Au$+$Au minimum bias, and Au$+$Au central events at $\sqrt{s_\mathrm{NN}}=200$~GeV \protect\cite{Jie}.}
\label{fig:sbratio}
\end{figure}

In the left and right panel of Fig.\ \ref{fig:backgrounds}, the unlike-sign, like-sign, and mixed-event distributions are shown for p$+$p and minimum bias Au$+$Au events, respectively. In p$+$p collisions, the like-sign method is used for $m_\mathrm{ee}<0.4$~GeV/$c^2$, and the mixed-event method for the higher mass region \cite{starpp}. In Au$+$Au collisions, the like-sign method is applied for $m_\mathrm{ee}<0.75$~GeV/$c^2$ and a combination of the mixed-event and the like-sign method for the higher mass region \cite{Jie}. The signal-to-background ratio is shown in Fig.\ \ref{fig:sbratio}.

\section{Dielectron Results}
In this section we present the recent STAR results on the dielectron production in p$+$p and Au$+$Au at $\sqrt{s_\mathrm{NN}}$=200~GeV. The p$+$p preliminary results are based on the data set that was taken in 2009. The results are based on 107~million events and includes 72\% of the TOF detector, which at that time was still in a commissioning phase. The Au$+$Au data was taken the next year, 2010, with the full TOF detector installed and operational. The 2010 data set involves a total of 270~million minimum-bias and 150~million central events.

\subsection{Dielectron spectra}
\begin{figure}[ht]
\centering
\includegraphics[width=0.45\textwidth,keepaspectratio]{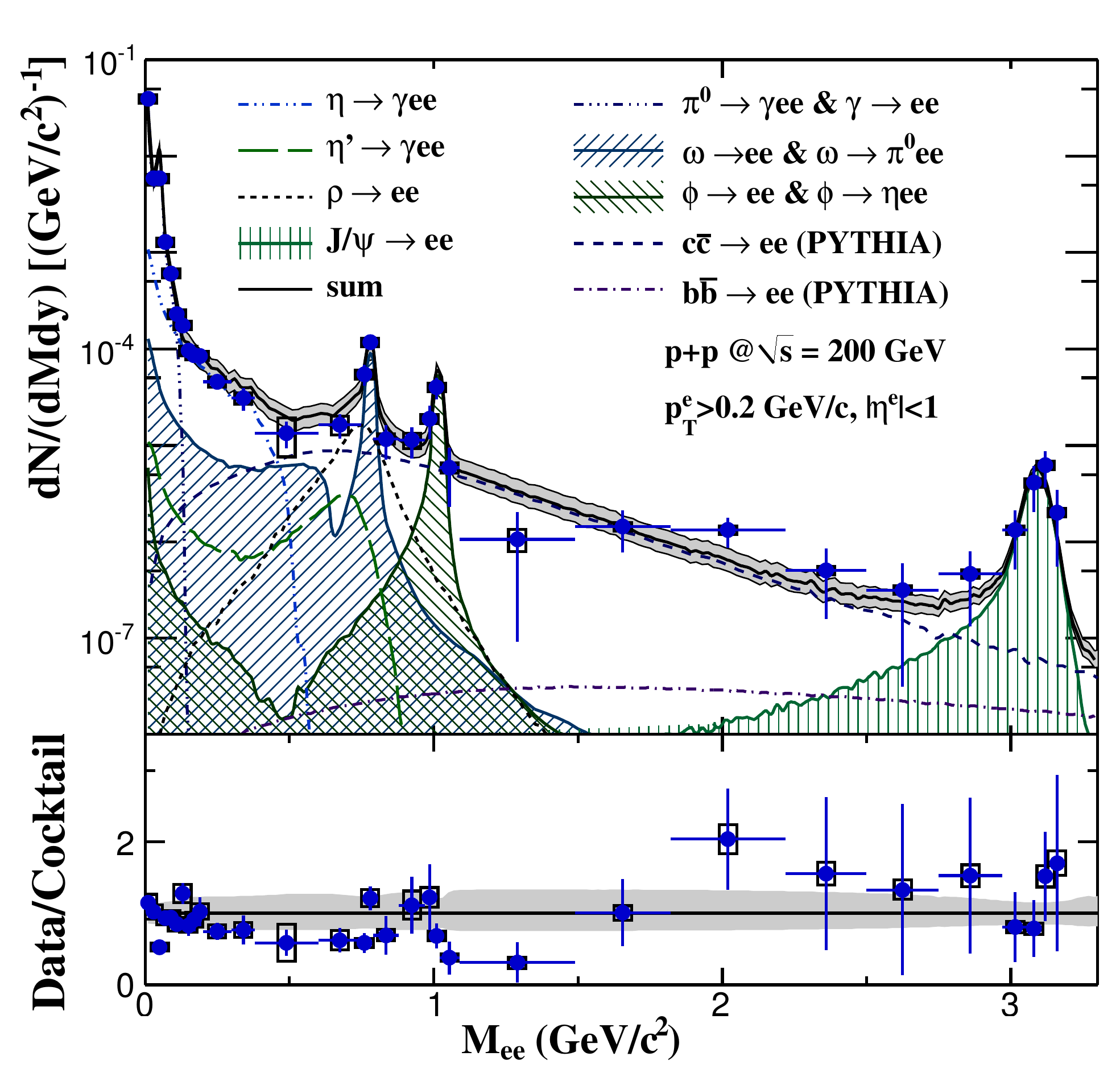}\\
\includegraphics[width=0.45\textwidth,keepaspectratio]{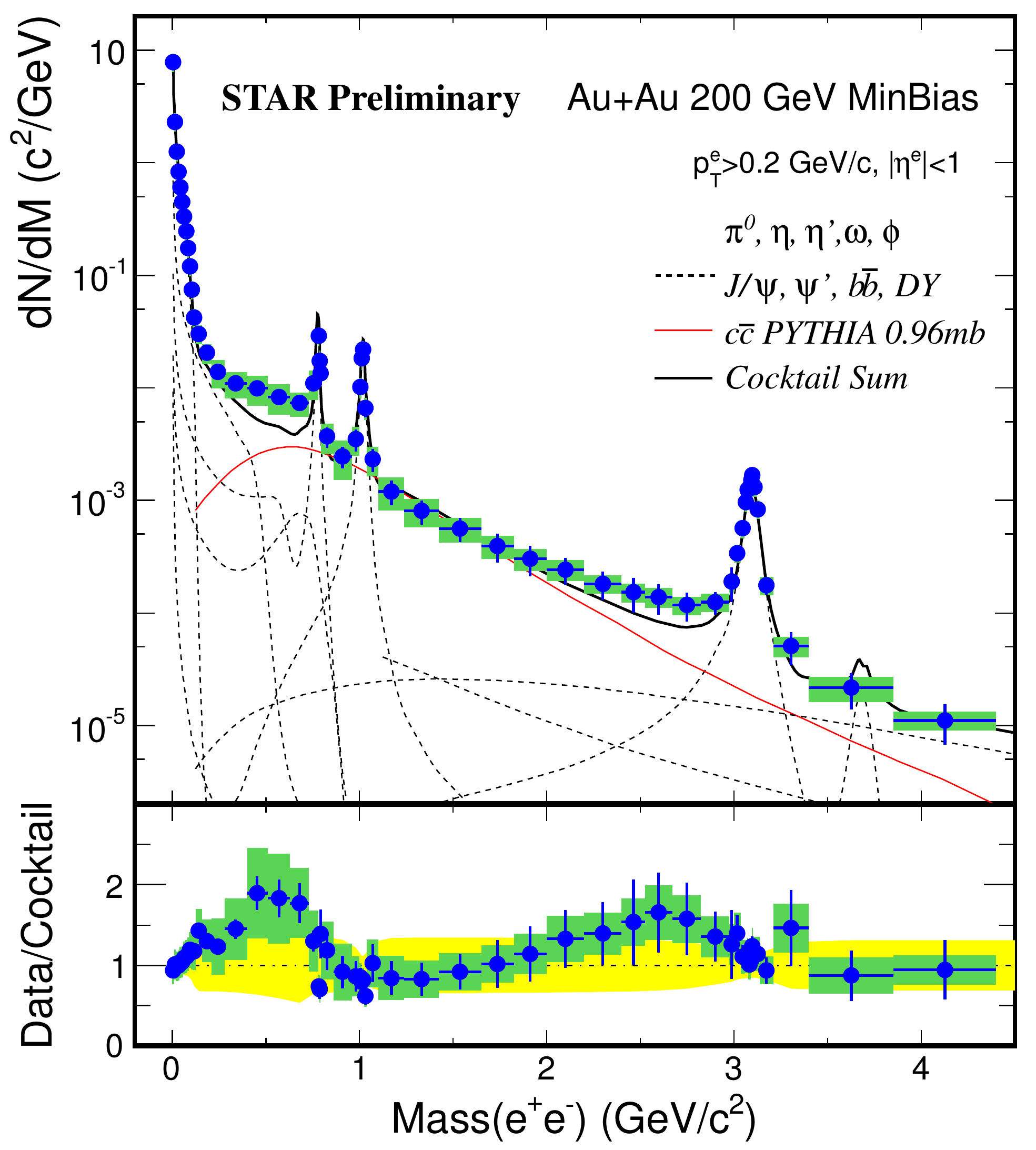}
\includegraphics[width=0.45\textwidth,keepaspectratio]{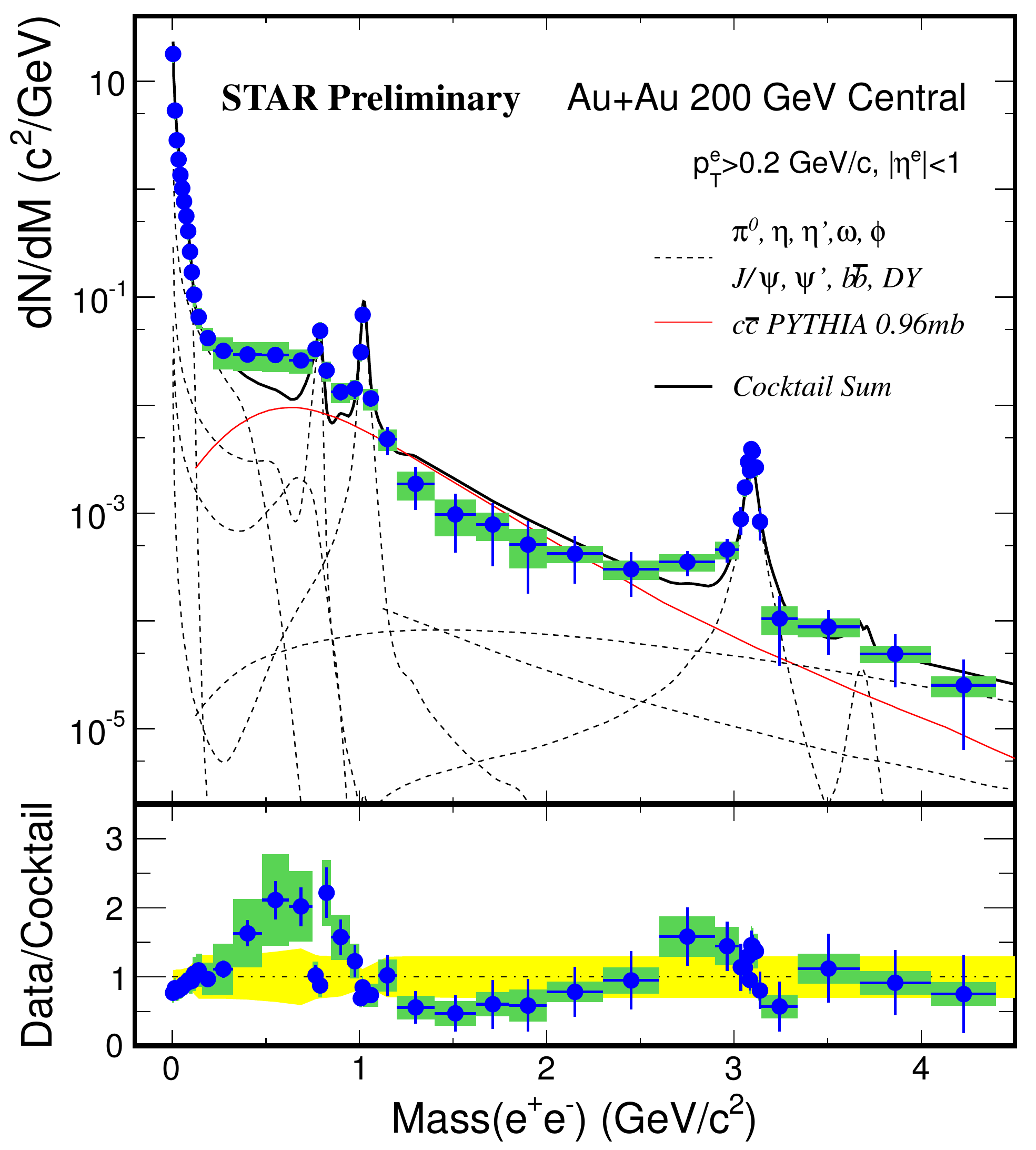}
\caption{Background subtracted invariant mass distributions for p$+$p \cite{starpp}  (upper), Au$+$Au minimum-bias (lower left), and Au$+$Au central collisions (lower right) at $\sqrt{s_\mathrm{NN}}$=200~GeV \protect\cite{Jie}. The grey and yellow bands are the systematic uncertainty on the cocktail simulation. The blue and green boxes indicate the systematic error on the data points.}
\label{fig:invmassspectra}
\end{figure}

 In Fig.\ \ref{fig:invmassspectra}, we show the dielectron LMR and IMR invariant mass spectra within the STAR acceptance ($|y_\mathrm{ee}| < 1.0, |\eta_\mathrm{e}|<1$, and $p_\mathrm{T}>0.2$~GeV/$c^2$) for p$+$p, Au$+$Au minimum-bias, and central collisions at $\sqrt{s_\mathrm{NN}}$=200~GeV. The spectra have been corrected for efficiency, but have not been corrected for radiation energy loss and momentum resolution.  The uncertainties on the data points include both statistical and systematic errors.
 
   The results are compared to a cocktail simulation of expected yields from several hadronic decays, heavy-flavor decays, and Drell-Yan production. The hadronic decays includes the leptonic decay channels of the $\omega$, $\rho^0$, $\phi$, and J/$\psi$ vector mesons, as well as the Dalitz decays of the $\pi^0$, $\eta$,  and $\eta'$ mesons. The input distributions to the simulations are based on Tsallis blast-wave function fits to the invariant yields of the measured mesons \cite{Lijuan}. These functions are used as the input distributions for the {\sc Geant} detector simulation using the full STAR geometry. In the IMR the dielectron production is dominated by the c$\bar{\mathrm{c}}$ cross section. This input, and the b$\bar{\mathrm{b}}$ cross section have been constrained by STAR measurements \cite{STARd0}.  More details including the simulation of the Dalitz decays can be found in \cite{Lijuan,Bingchu}. For the   Au$+$Au spectra, the $\rho^0$ contribution has not been included in the cocktail as it may be sensitive to in-medium modifications possibly affecting its mass and spectral line shape \cite{starrho0}. The individual contributions to the cocktail as well as the sum of all contributions are shown in the upper subpanels of  Fig.\  \ref{fig:invmassspectra}. The lower subpanels show the ratio of the measured dielectron spectra to the cocktail. Uncertainties on both the data and cocktail are included.
  
The upper panel of Fig.\ \ref{fig:invmassspectra} shows the data and cocktail simulations for the  p$+$p collisions. The data is consistent with the cocktail and will provide a baseline for the Au$+$Au systems. In the lower left panel of Fig.\ \ref{fig:invmassspectra}, the data from the minimum bias Au$+$Au collisions are shown. In the cocktail simulations, the c$\bar{\mathrm{c}}$$\rightarrow$e$^+$e$^-$ cross section is based on {\sc Pythia} simulations scaled by the number of binary nucleon-nucleon collisions \cite{Jie}. We observe a hint of enhancement in the LMR.  In comparison with the minimum-bias data, we observe this enhancement to be more pronounced in central collisions, as can be seen in the right panel of Fig.\ \ref{fig:invmassspectra}. To further quantify the observed enhancement, we report the enhancement factor, {\em i.e.} the ratio of data to cocktail, in the range of  $150 < m_\mathrm{ee} < 750$~MeV/$c^2$. We find the enhancement factors to be 1.53$\pm$0.07$^\mathrm{stat}$$\pm$0.41$^\mathrm{syst}$ and 1.72$\pm$0.10$^\mathrm{stat}$$\pm$0.50$^\mathrm{syst}$ for minimum-bias and central events, respectively.  These enhancement factors are based on cocktail simulations that do not include contributions from the $\rho^0$ meson. The LMR  enhancement does not exclude possible in-medium modifications of the vector mesons and differential measurements of centrality and transverse momentum need to be further studied.
Measurements from both the Pioneering High Energy Nuclear Interaction Experiment (PHENIX) \cite{phenix} and the STAR experiment indicate an LMR enhancement. However, a quantitative comparison of the published PHENIX enhancement factors with the STAR preliminary results shows a significant difference:  4.7$\pm$0.04$^\mathrm{stat}$$\pm$1.5$^\mathrm{syst}$$\pm$0.9$^\mathrm{model}$ and 7.6$\pm$0.05$^\mathrm{stat}$$\pm$1.3$^\mathrm{syst}$$\pm$1.5$^\mathrm{model}$ for minimum bias and top-10\% central events, respectively. Further studies which carefully take acceptance differences into account are needed to help understand this discrepancy.

In the IMR for central Au$+$Au collisions, we observe the cocktail simulation to slightly overshoot the data and while still consistent within uncertainties this may hint at an in-medium modification of the c$\bar{\mathrm{c}}$ production. With the complement of future detector upgrades, STAR will be in an excellent position to further study such modifications, as will be discussed in Sect. \ref{sect:prospect}.

\begin{figure}[ht]
\centering
\includegraphics[width=0.48\textwidth,keepaspectratio]{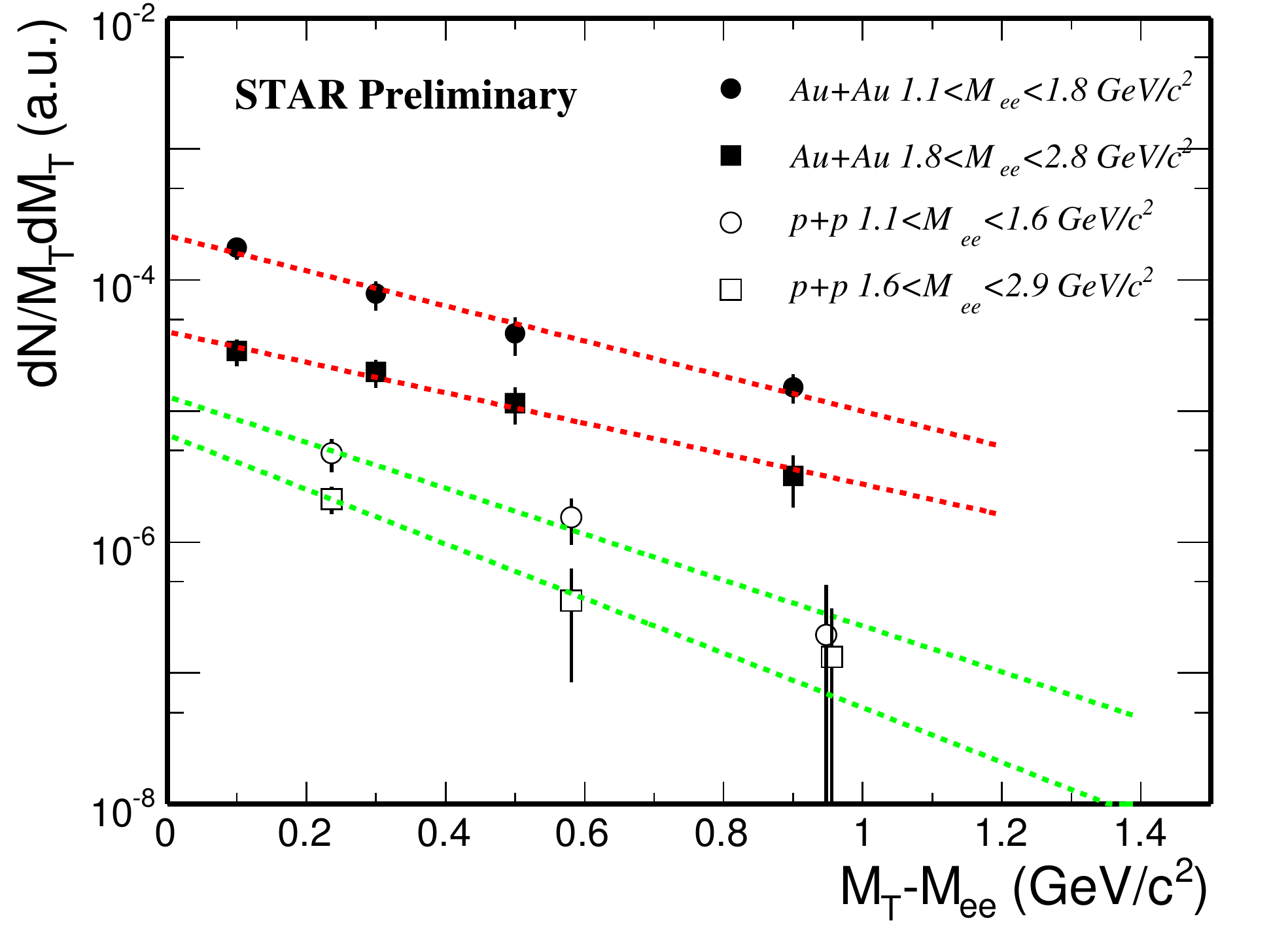}
\includegraphics[width=0.48\textwidth,keepaspectratio]{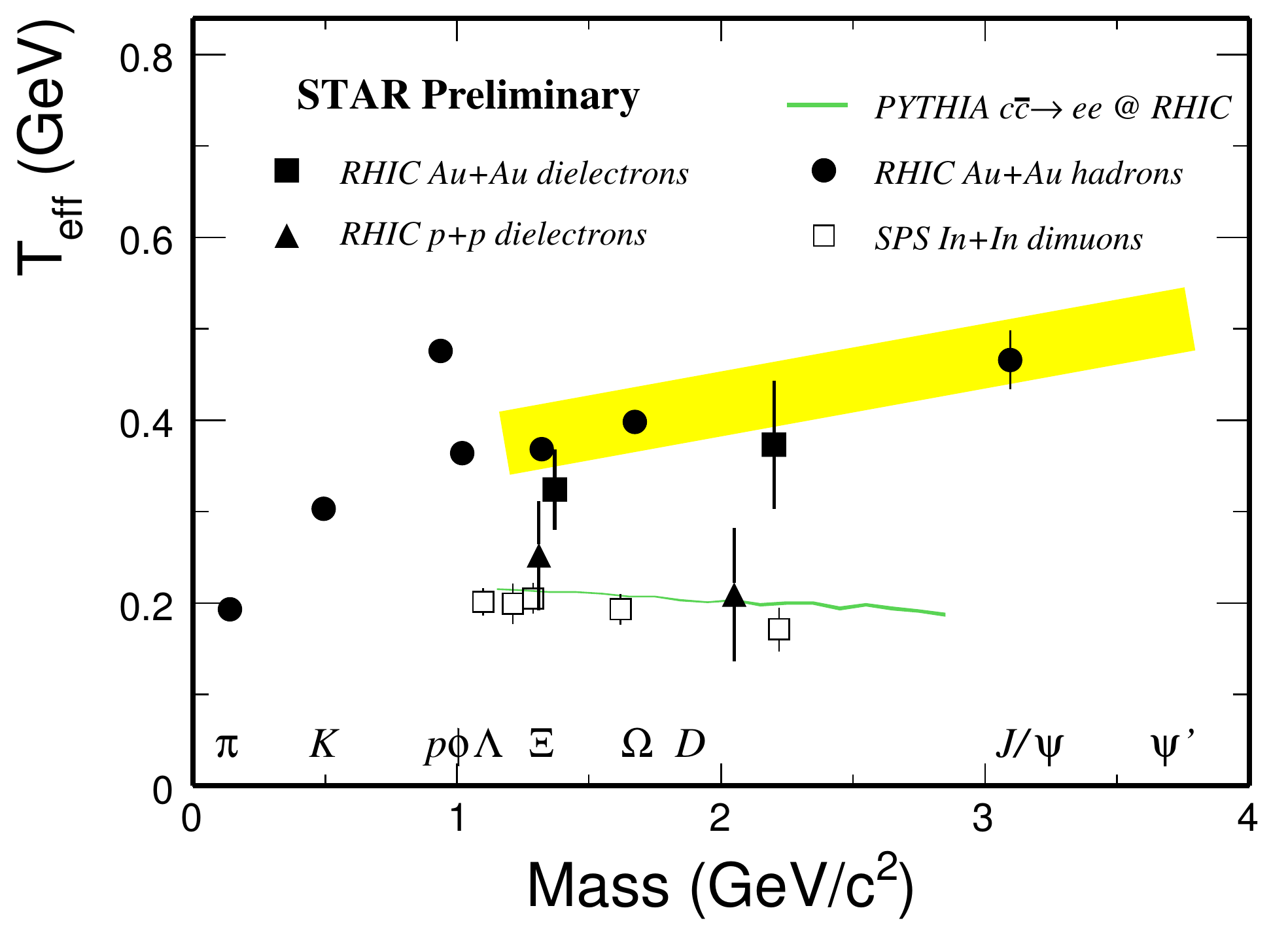}
\caption{Left panel: Transverse mass spectra for dielectrons in two invariant mass ranges, for  p$+$p and minimum bias Au$+$Au at $\sqrt{s_\mathrm{NN}}$=200~GeV. Right panel: Transverse mass slope parameters from RHIC and SPS measurements \protect\cite{Jie}. Uncertainties are statistical only. }
\label{fig:dielectron_mt}
\end{figure}

In the left panel of Fig.\ \ref{fig:dielectron_mt}, the transverse mass distributions, $m_\mathrm{T}-m_\mathrm{ee}$, for two IMR invariant mass ranges are shown for the p$+$p and minimum bias Au$+$Au systems. In the right panel of Fig.\ \ref{fig:dielectron_mt}, we observe that the transverse mass slope parameters in p$+$p collisions are consistent with {\sc Pythia} simulations. The Au$+$Au results  are higher than the p$+$p slope parameters. This may be due to in-medium charm modification or contributions from QGP thermal radiation. In the same panel of Fig.\ \ref{fig:dielectron_mt}, we compare with the SPS-energy dimuon results from the NA60 experiment \cite{na60}. The NA60 results have contributions from open charm decay and Drell-Yan pairs subtracted. The inclusive transverse mass slopes for dielectrons measured by STAR at top RHIC energies in Au$+$Au collisions are observed to be larger than what has been measured for the SPS-energy dimuons.

\subsection{Dielectron flow}

\begin{figure}[h]
\centering
\includegraphics[width=0.45\textwidth,keepaspectratio]{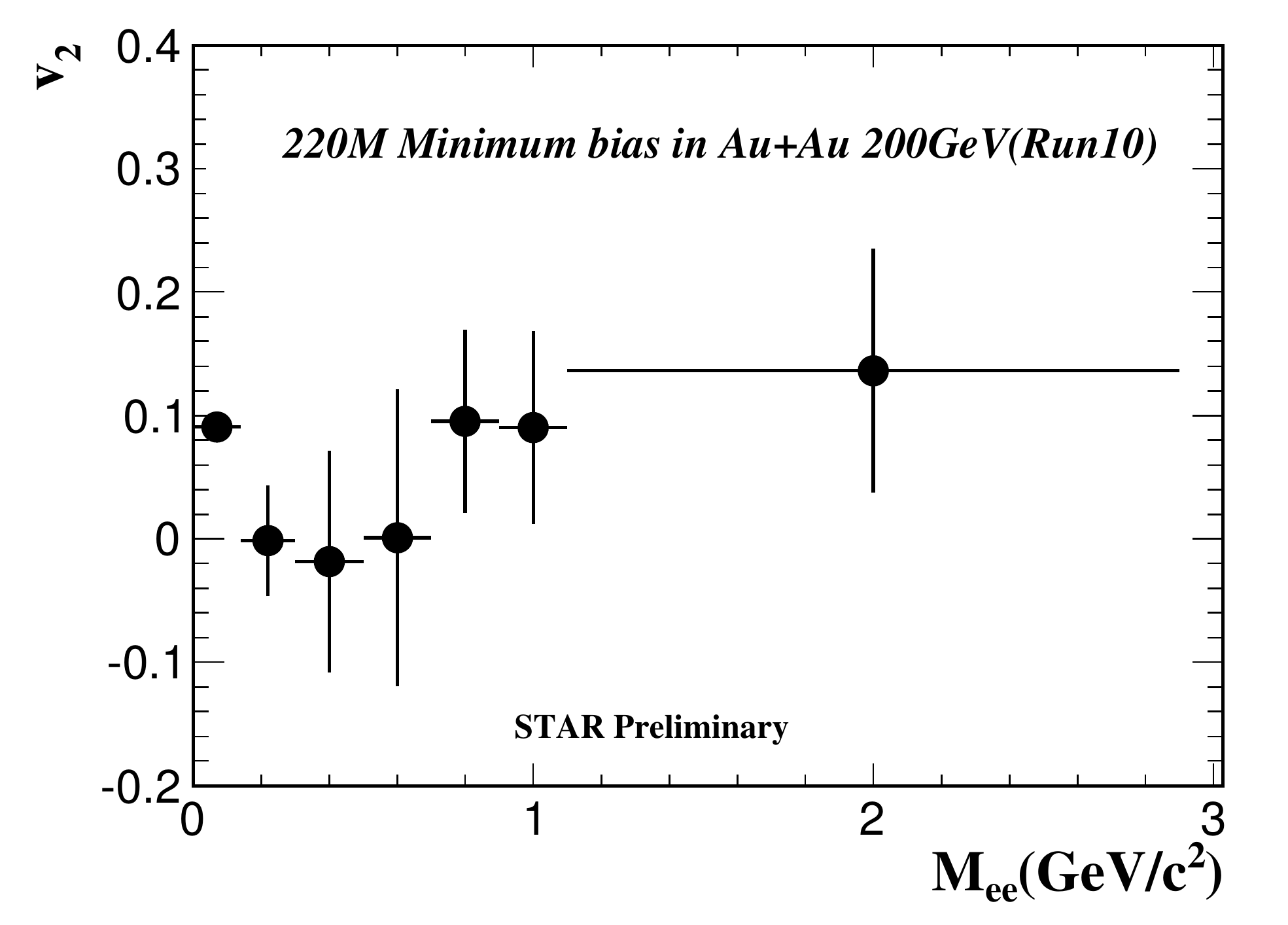}
\caption{Elliptic flow as a function of the dielectron invariant mass in minimum-bias Au$+$Au collisions at $\sqrt{s_\mathrm{NN}}$=200~GeV. Uncertainties are statistical only.}
\label{fig:flowmass}
\end{figure}

 In Fig.\ \ref{fig:flowmass}, we present the first STAR dielectron elliptic-flow measurements as a function of the dielectron mass. The results are based on 220~million Au$+$Au minimum bias events at $\sqrt{s_\mathrm{NN}}$=200~GeV.  The elliptic flow, $v_2$, is calculated using the event-plane method in which the event plane is reconstructed from TPC tracks \cite{hadronflow}.
 The ``signal" elliptic flow, $v_2^\mathrm{signal}$, is calculated as follows \cite{sqm11}:
\[
 v_2^\mathrm{total} (m_\mathrm{ee}) = v_2^\mathrm{sign} (m_\mathrm{ee}) \frac{r(m_\mathrm{ee})}{1+r(m_\mathrm{ee})}  +  v_2^\mathrm{bkgd} (m_\mathrm{ee})\frac{1}{1+r(m_\mathrm{ee})},
\]
where $ v_2^\mathrm{total}$ is the flow measurement of all dielectron candidates, $v_2^\mathrm{bkgd}$ the flow measurement on the dielectron background, and $r(m_\mathrm{ee})$ the mass-dependent signal-to-background ratio, shown in Fig.\ \ref{fig:sbratio}.

\begin{figure}[h]
\centering
\includegraphics[width=0.45\textwidth,keepaspectratio]{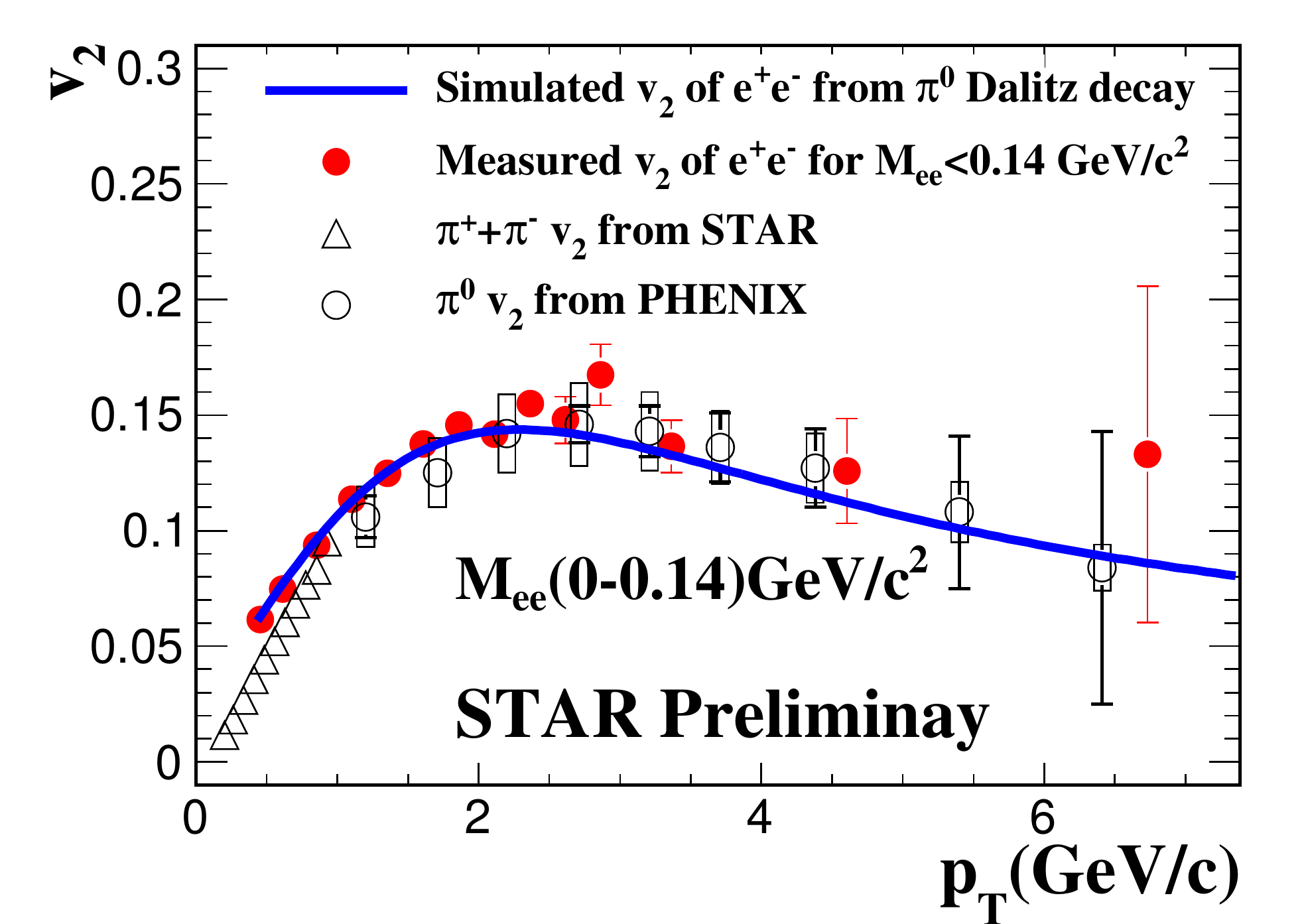}
\includegraphics[width=0.45\textwidth,keepaspectratio]{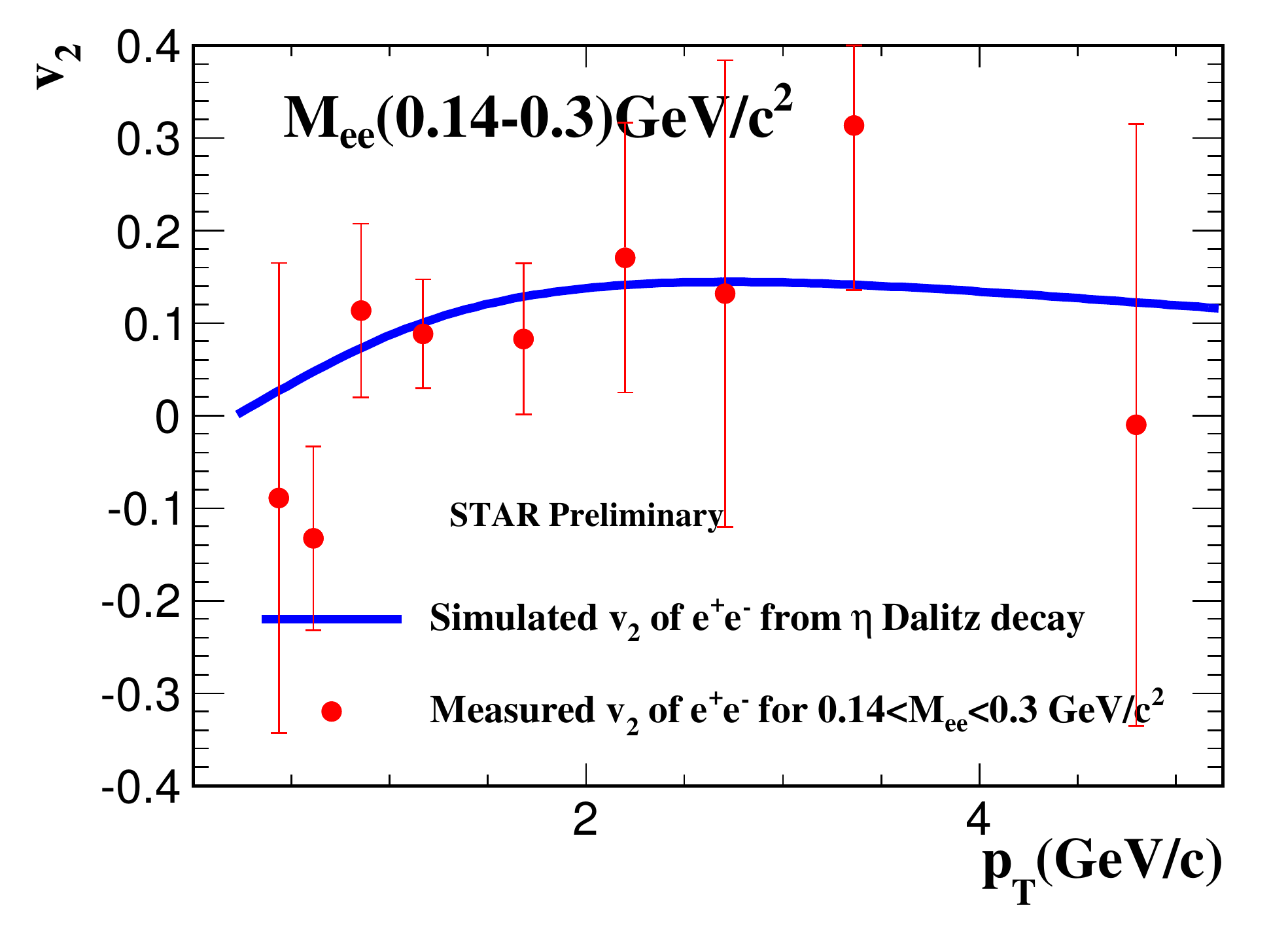}
\caption{Left panel: The elliptic flow as a function of transverse momentum for two dielectron invariant mass regions in Au$+$Au at $\sqrt{s_\mathrm{NN}}$=200~GeV (red symbols). The $\pi^0$ (open circle) and $\pi^\pm$ (triangle) elliptic flow measurements, from \protect\cite{phenixpi0,starpiv2}, respectively, and the expected elliptic flow of dielectrons from $\pi^0$ Dalitz decays (solid line). Right panel: The elliptic flow of the dielectron in the $\eta$ Dalitz decay region as a function of transverse momentum in minimum-bias Au$+$Au collisions at $\sqrt{s_\mathrm{NN}}$=200~GeV (red symbols). The solid line shows the expected elliptic flow of dielectrons from $\eta$ Dalitz decays. Uncertainties are statistical only.}
\label{fig:flowmasspid}
\end{figure}

In Fig.\ \ref{fig:flowmasspid},  the $v_2$ differential as a function of $p_\mathrm{T}$ is shown for the two dielectron mass ranges that correspond to the $\pi^0$ and $\eta$ Dalitz decay ranges. The left panel shows $v_2$ for  the $\pi^0$ Dalitz decay region, $m_\mathrm{ee}<$140~MeV/$c^2$. The measurements are compared with the $\pi^0$ $v_2$ measured by the PHENIX Collaboration \cite{phenixpi0} and the $\pi^\pm$ $v_2$ measured by STAR \cite{starpiv2}. Furthermore, the published pion flow measurements have been parametrized and used as input for a $\pi^0$ Dalitz decay simulation to obtain the anticipated dielectron $v_2$. We observe that the dielectron $v_2$ measurements agree very well \cite{moriond2011}.

A similar procedure is applied to compare the measured $v_2$ in the $\eta$ Dalitz-decay range, 140$<m_\mathrm{ee}<$300~MeV/$c^2$, with the expected $v_2$ of the Dalitz-decayed $\eta$ meson. In this case, however, the input distribution is based on the published STAR K$_\mathrm{S}^0$ $v_2$ measurements \cite{hadronflow} since the masses of both mesons are comparable. We observe that in this range the $v_2$ data and simulations are consistent. The agreement between measurement and expectation in both the $\pi^0$ and $\eta$ Dalitz-decay regions supports the validity of the methods that we use to reconstruct the dielectron $v_2$. 

\section{Summary \& Prospects}
\label{sect:prospect}
 We have presented the dielectron results for p$+$p and Au$+$Au collisions at $\sqrt{s_\mathrm{NN}}=$200~GeV. When compared to a hadron cocktail, we observe the p$+$p data to be consistent with this cocktail and note that the charm contribution dominates in the IMR. In minimum-bias Au$+$Au collisions, when compared to the hadron cocktail, we observe a hint of an enhanced production in the LMR. This enhancement is more pronounced in central  Au$+$Au collisions. We have also reported on the first STAR dielectron elliptic flow measurements. A comparison of the observed $v_2$ in the $\pi^0$ and $\eta$ Dalitz decay regions with other elliptic flow measurements shows reasonable consistency.
 
 \begin{figure}[ht]
\centering
\includegraphics[width=0.45\textwidth,keepaspectratio]{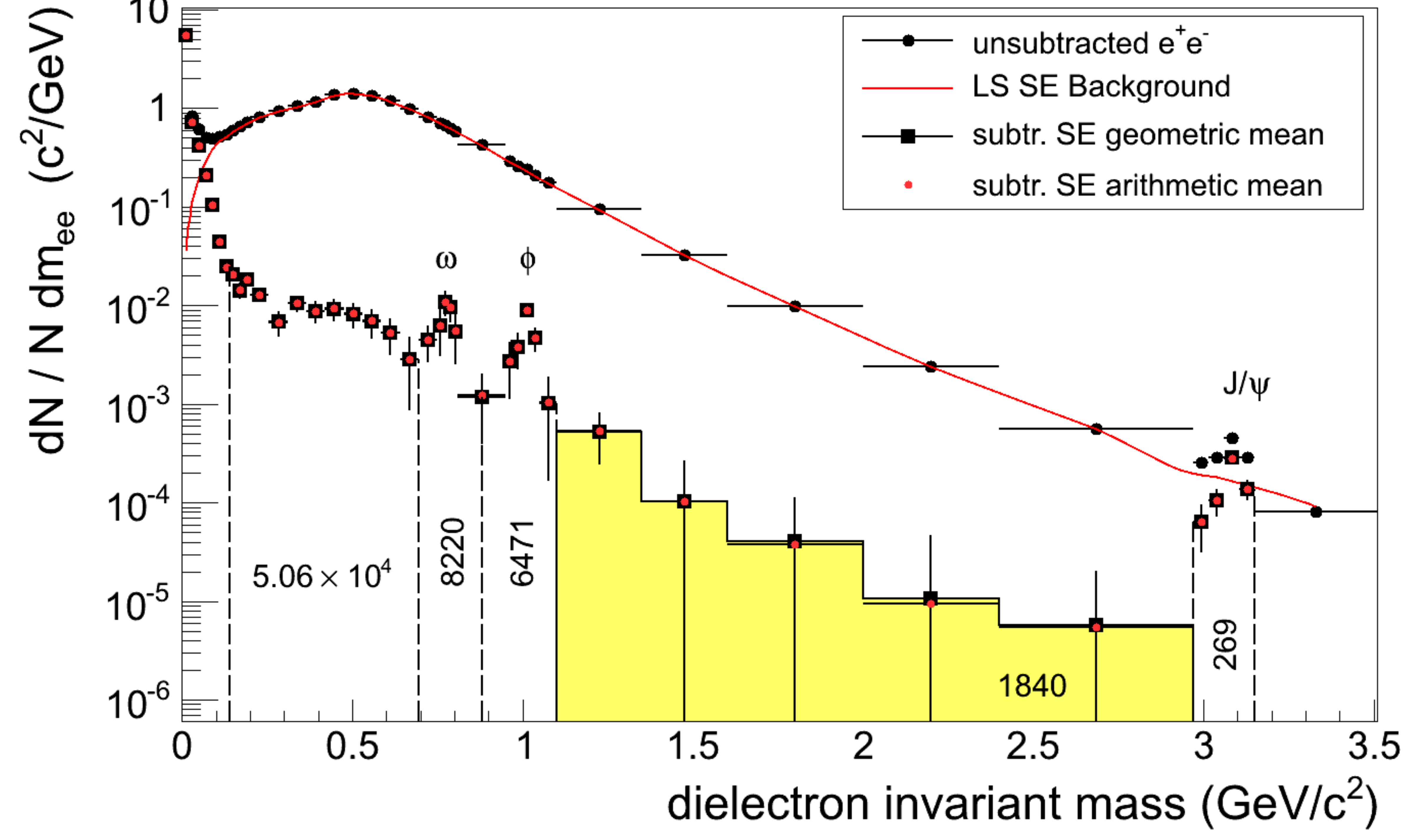}
\caption{Raw and background-subtracted dielectron invariant mass distributions for minimum-bias Au$+$Au collisions at $\sqrt{s_\mathrm{NN}}$=39~GeV. Uncertainties are statistical only. }
\label{fig:raw39GeV}
\end{figure}
 
The results presented in this paper have all been based on data taken in 2009 (p$+$p) and 2010 (Au$+$Au). Recently, the STAR experiment significantly increased the statistics for both the p$+$p (2012) and Au$+$Au (2011) data samples. In addition to these samples at the RHIC top energy, the STAR experiment has collected data at the lower beam energies of $\sqrt{s_\mathrm{NN}}=19.6 - 62$GeV in 2010 and 2011.
 At these energies, the dielectron analyses benefit from a fully installed TOF detector as well as a low conversion material budget and systematically study the energy dependence of the pair production in the LMR and IMR ranges.  In Fig.\ \ref{fig:raw39GeV}, we show the raw and background-subtracted dielectron mass distributions, based on 110~million minimum bias Au$+$Au collisions at $\sqrt{s_\mathrm{NN}}=39$~GeV, taken by the STAR experiment in 2010 \cite{Patrick}.  The study of dielectron production at lower energies will allow us to address the beam-energy dependence of the observed LMR dielectron enhancement, the in-medium modifications of the vector mesons, as well as the IMR medium modifications to the charm decay, and the possible QGP thermal radiation.

In the intermediate invariant mass range, the c$\bar{\mathrm{c}}$ medium modifications and QGP thermal radiation are difficult to disentangle. The upcoming Heavy Flavor Tracker upgrade \cite{hft} will provide very useful tools that will allows us to measure the charm cross sections more precisely, but the  c$\bar{\mathrm{c}}$  measurement may still be very challenging. Alternatively, the MTD upgrade \cite{mtd} will allow us to measure e$+$$\mu$ correlations which measure the contributions from heavy-flavor decays to the dilepton continuum. This should improve our access to the QGP thermal radiation in this region. The HFT and MTD upgrade projects are scheduled for completion in 2014.

\section*{Acknowledgements}
This work is supported by the US Department of Energy under grant DE-FG02-10ER41666.


\begin{thebibliography}{}
\bibitem{whitepaperBRAHMS} BRAHMS Collaboration Nucl.\ Phys.\ A \textbf{757} (2005), 1.
\bibitem{whitepaperPHOBOS} PHOBOS Collaboration Nucl.\ Phys.\ A \textbf{757} (2005), 28.
\bibitem{whitepaperSTAR} STAR Collaboration, Nucl.\ Phys.\ A \textbf{757} (2005), 102.
\bibitem{whitepaperPHENIX} PHENIX Collaboration, Nucl.\ Phys.\ A \textbf{757} (2005), 184.
\bibitem{Chatterjee}  R.~Chatterjee {\em et al.}, Phys.\ Rev.\ C \textbf{75} (2007), 054909.
\bibitem{llope} W.~Llope (for the STAR Collaboration), Nucl.\ Instr. and Meth.\ A \textbf{661} (2012), S110.
\bibitem{mtd} L.~Ruan {\em et al.}, J.\ Phys.\ G \textbf{36} (2009), 095001.
\bibitem{Lijuan} L.~Ruan (for the STAR Collaboration),Nucl.\ Phys.\ A \textbf{855} (2011), 269.
\bibitem{starpp} L.~Adamczyk {\em et al.}, arXiv:1204.1890.
\bibitem{Jie} J.~Zhao (for the STAR Collaboration), J.\ Phys.\ G: Nucl.\ Part.\ Phys.\ \textbf{38} (2011), 124134.
\bibitem{STARd0} J.~Adams {\em et al.}, Phys.\ Rev.\ Lett.\ \textbf{94} (2005), 062301.
\bibitem{Bingchu} B.~Huang, \textit{Di-lepton production measurements in p+p and Au+Au collisions at RHIC}, Ph.D.\ Thesis, USTC, Hefei, China (2011).
\bibitem{starrho0} J.~Adams {\em et al.}, Phys.\ Rev.\ Lett.\ \textbf{92} (2004), 092301.
\bibitem{phenix} A.~Adare {\em et al.}, Phys.\ Rev.\ C \textbf{81} (2010), 034911.
\bibitem{na60} R.~Arnaldi {\em et al.},  Phys.\ Rev.\ Lett.\ \textbf{100}  (2008), 022302.
\bibitem{hadronflow} B.~Abelev {\em et al}, Phys.\ Rev.\ C \textbf{77} (2008), 054901.
\bibitem{sqm11} B.~Huang (for the STAR Collaboration), Acta Phys.\ Pol.\ B Proc.\ \textbf{5} (2012), 471.
\bibitem{phenixpi0} S.~Afanasiev {\em et al}, Phys.\ Rev.\ C \textbf{80} (2009), 054907.
\bibitem{starpiv2} Y.~Bai, \textit{Anisotropic Flow Measurements in STAR at the Relativistic Heavy Ion Collider}, Ph.D.\ Thesis, Universiteit Utrecht, The Netherlands (2007).
\bibitem{moriond2011} L.~Ruan  (for the STAR Collaboration), in Proc.\  of the 47$^\mathrm{th}$ Recontres de Moriond (2012)
\bibitem{Patrick} P.~Huck (for the STAR Collaboration), CPOD Workshop (2011), CCNU Wuhan.
\bibitem{hft} S.~Kleinfelder {\em et al.}, Nucl.\ Instr. and Meth. A \textbf{565} (2006), 132.
\end{thebibliography}
\end{document}